\documentclass[11pt]{article}
\pdfoutput=1
\usepackage{geometry}   
\usepackage{graphicx}
\usepackage{amsfonts}
\usepackage{amsmath}
\usepackage{natbib}
\usepackage{ctable}
\usepackage{lscape}
\usepackage{multirow}

\title{Bayesian Sparse Factor Analysis of Genetic Covariance Matrices  \\
}

\author{Daniel E. Runcie\thanks{Department of Biology, Duke
    University, Durham, NC 27708}, Sayan Mukherjee\thanks{Departments of
    Statistical Science, Computer Science, and Mathematics, Duke 
    University, Durham, NC 27708}}

\date{}

\newcommand{\bm}{\mathbf}

\renewcommand{\cite}{\citep}
\newcommand{\citeN}{\citet}
\newcommand{\citeNP}{\citealt}

\begin{document}

\maketitle

\begin{abstract}
Quantitative genetic studies that model complex, multivariate
phenotypes are important for both evolutionary prediction and
artificial selection. For example, changes in gene expression can
provide insight into developmental and physiological mechanisms that
link genotype and phenotype. However, classical analytical techniques
are poorly suited to quantitative genetic studies of gene expression
where the number of traits assayed per individual can reach many
thousand. Here, we derive a Bayesian genetic sparse factor model for
estimating the genetic covariance matrix (G-matrix) of
high-dimensional traits, such as gene expression, in a mixed effects
model. The key idea of our model is that we need only consider
G-matrices that are biologically plausible. An organism's entire
phenotype is the result of processes that are modular and have limited
complexity. This implies that the G-matrix will be highly
structured. In particular, we assume that a limited number of
intermediate traits (or factors, e.g., variations in development or
physiology) control the variation in the high-dimensional phenotype,
and that each of these intermediate traits is sparse -- affecting only
a few observed traits. The advantages of this approach are
two-fold. First, sparse factors are interpretable and provide
biological insight into mechanisms underlying the genetic
architecture. Second, enforcing sparsity helps prevent sampling errors
from swamping out the true signal in high-dimensional data. We demonstrate the advantages of our model on simulated data and in an analysis of a published \textit{Drosophila melanogaster} gene expression data set.

\end{abstract}

\section{Introduction}
	
Quantitative studies of evolution or artificial selection often focus on a single or a handful of traits at a time, such as size, survival or crop yield. Recently, there has been an effort to collect more comprehensive phenotypic information on traits such as morphology, behavior, physiology, or gene expression \cite{Houle:2010jw}. For example, the expression of thousands of genes can be measured simultaneously \cite{Ayroles:2009gd,Mcgraw:2011dm,Gibson:2005tw}, together capturing complex patterns of gene regulation that reflect molecular networks, cellular stresses, and disease states \cite{xiong11,delacruz10}, and may in some cases be important for fitness. Studying the quantitative genetics  of multiple correlated traits requires a joint modeling approach  \cite{Walsh:2009gg}. However, applying the tools of quantitative genetics to high-dimensional, highly correlated datasets presents considerable analytical and computational challenges \cite{Meyer:2010jj}. In this paper we formulate a modeling framework to address these challenges for a basic component of quantitative genetic analysis: estimation of the matrix of additive genetic variances and covariances, or G-matrix \cite{Lynch:1998vx}. The G-matrix encodes information about responses to selection \cite{Lande:1979th}, evolutionary constraints \cite{Kirkpatrick:2009er}, and modularity \cite{Cheverud:1996dc}, and is important for predicting evolutionary change \cite{Schluter:1996up}. Thus, G-matrix estimation is a key step for many quantitative genetic analyses.

The challenge in scaling classic methods to hundreds or thousands of
traits is that the number of modeling parameters grows
exponentially. An unconstrained G-matrix for $p$ traits requires
$p(p+1)/2$ parameters, and modeling environmental variation and
measurement error \cite{Kirkpatrick:2004bv} requires at least as many
additional parameters. Coupled with modest sample sizes, huge numbers
of parameters can lead to instability in parameter estimates --
analyses that are highly sensitive to outliers and have high
variance. Previous methods for overcoming this instability include (1)
``bending" or smoothing unconstrained estimates of G-matrices, such as
from pairwise estimates of genetic covariation
\cite{Ayroles:2009gd,Stone:2009jx} or moments estimators
\cite{Hayes:1981uc}, and (2) estimating a constrained G-matrix to be
low rank and thus specified with fewer parameters (e.g.,
\citeNP{Kirkpatrick:2004bv}). Constraining the G-matrix has
computational and analytical advantages: fewer parameters result in
more robust estimates and lower computational requirements
\cite{Kirkpatrick:2004bv}. Constrained estimators of G-matrices
include methods based on moments estimators
\cite{Hine:2006ki,Mcgraw:2011dm} and methods based on mixed effects
models (e.g., the ``animal model" and other related models
\cite{Henderson:1984vh,Kruuk:2004be,Kirkpatrick:2004bv,deLosCampos:2007ib}. 
Mixed effects and related models have been particularly powerful for studies in 
large breeding programs and wild populations. These methods perform
well on moderate-dimensional data. However, they are too
computationally costly and not sufficiently robust to analyze 
high-dimensional traits. 

Our objective in this paper is to develop a model for estimating
G-matrices that is scalable to large numbers of traits and is
applicable to a variety of experimental designs, including both
experimental crosses and pedigreed populations. We build on the
Bayesian mixed effects model of \citeN{deLosCampos:2007ib} and model
the G-matrix with a factor model, but add additional constraints by
using a highly informative, biologically-motivated, prior
distribution. The key idea that allows us to scale to large numbers of
traits is that the vast majority of the space of covariance matrices does not contain matrices that are biologically plausible as a G-matrix: we expect the G-matrix to be \emph{sparse}, by which we mean that we favor G-matrices that are \emph{modular} and \emph{low-rank}. Sparsity in statistics refers to models in which many parameters are expected to be zero \cite{lucas2006sparse}. By modular, we mean that small groups of traits will covary together. By low-rank, we mean that there will be few (important) modules. We call a G-matrix with these properties \emph{sparse} because there exists a low-rank factorization (most of the possible dimensions are zero) of the matrix with many of its values equal to (or close to) zero. This constrains the class of covariance matrices that we search over, a necessary procedure for inference of covariance matrices from high-dimensional data \cite{bickel1,bickel2,elkaroui,Meyer:2010jj,Carvalhoetal2008,Hahnetal2013}. Under these assumptions, we can also interpret the modules underlying our factorization without imposing additional constraints such as orthogonality \cite{englehardt}, something not possible with earlier mixed effect factor models \cite{Meyer:2009jm}.

The biological argument behind this prior assumption starts with the observation that the observed traits of an organism arise from common developmental processes of limited complexity, and developmental processes are often modular \cite{Cheverud:1996dc,Wagner:1996tq,Davidson:2008ui}. For gene expression, regulatory networks control gene expression, and variation in gene expression can be often linked to variation in pathways \cite{xiong11,delacruz10}. For a given dataset, we make two assumptions about the modules (pathways): (1) a limited number of modules contribute to trait variation and (2) each module affects a limited number of traits. There is support and evidence for these modeling assumptions in the quantitative genetics literature as G-matrices tend to be highly structured \cite{Walsh:2009gg} and the majority of genetic variation is contained in a few dimensions regardless of the number of traits studied \cite{Ayroles:2009gd,Mcgraw:2011dm}. Note that while we focus on developmental mechanisms underlying trait covariation, ecological or physiological processes can also lead to modularity in observed traits and our prior may be applied to these situations as well. 

Based on these assumptions, we present a Bayesian sparse factor model for inferring G-matrices from pedigree information for hundreds or thousands of traits. We demonstrate the advantages of the model on simulated data and re-analyze gene expression data from a published study on \textit{Drosophila melanogaster} \cite{Ayroles:2009gd}. Although high-dimensional sparse models have been widely used in genetic association studies \cite{cantoretal,englehardt,Stegle:2010cg,Parts:2011df,Zhouetal2012} to our knowledge, sparsity has not yet been applied to estimating a G-matrix. 
	
\section{Methods}

In this section, we derive the Bayesian genetic sparse factor model as an extension to the classic multivariate animal model to the high-dimensional setting, where hundreds or thousands of traits are simultaneously examined. A factor model posits that a set of unobserved (latent) traits called \emph{factors} underly the variation in the observed (measured) traits. For example, measured gene expression traits might be the downstream output of a gene regulatory network. Here, the activity of this gene network is a latent trait which might vary among individuals. We use the animal model framework to partition variation in both the measured traits and the latent factor traits into additive genetic variation and residuals. We encode our two main biological assumptions on the G-matrix as priors on the factors: sparsity in the number of factors that are important, and sparsity in the number of measured traits related to each factor. These priors constrain our estimation to realistic matrices and thus prevent sampling errors swamping out the true signal in high-dimensional data.

\subsection{Model:}

For a single trait the following linear mixed effects model is commonly used to explain phenotypic variation \cite{Henderson:1984vh}: 

\begin{align}
\label{single_trait_mixedModel}
\mathbf{y}_i = \mathbf{X}\mathbf{b}_i + \mathbf{Z}\mathbf{u}_i +  \mathbf{e}_i,
\end{align} 
\noindent where $\mathbf{y}_i$ is the vector of phenotype measurements
for trait $i$ on $n$ individuals; $\mathbf{b}_i$ is a vector of
coefficients for the fixed effects and environmental covariates such
as sex or age with design matrix $\mathbf{X}$; $\mathbf{u}_i \sim
\mbox{N}(\mathbf{0},\sigma^2_{G_i}\mathbf{A})$ is the random vector of
additive genetic effects with incidence matrix $\mathbf{Z}$,  and
$\mathbf{e}_i \sim \mbox{N}(\mathbf{0},\sigma^2_{R_i}\mathbf{I}_n)$ is
the residual error caused by non-additive genetic variation, random
environmental effects, and measurement error. The residuals are
assumed to be independent of the additive genetic effects. Here,
$\mathbf{A}$ is the known $r \times r$ additive relationship matrix
among the individuals; $r$ generally equals $n$, but will not if there are unmeasured parents, or if several individuals are clones and share the same genetic background (e.g., see the Drosophila gene expression data below).

In going from one trait to $p$ traits we can align the vectors $\mathbf{y}_i$ for each trait in
\eqref{single_trait_mixedModel} to form an $n \times p$ matrix $\mathbf{Y}$ specified 
by the following multivariate model:

\begin{align}
\label{multi_trait_mixedModel}
\mathbf{Y} = \mathbf{XB} + \mathbf{ZU} + \mathbf{E},
\end{align}
where $\mathbf{B} = [\mathbf{b}_1 \dots \mathbf{b}_p]$. $\mathbf{U} = [\mathbf{u}_1 \dots \mathbf{u}_p]$ and $\mathbf{E} = [\mathbf{e}_1 \dots \mathbf{e}_p]$ are random variables drawn from matrix normal distributions \cite{Dawid:1981gb}:
\begin{equation}
\label{random_effect_distributions}
\mathbf{U} \sim \mbox{MN}_{r,p}(\mathbf{0};\mathbf{A},\mathbf{G}), \quad
\mathbf{E} \sim \mbox{MN}_{n,p}(\mathbf{0};\mathbf{I}_n,\mathbf{R}),
\end{equation}
where the subscripts $r,p$ and $n,p$ specify the dimensions of the matrices, $\mathbf{0}$ is a matrix of zeros of appropriate size, $\mathbf{A}$ and $\mathbf{I}_n$ specify the row (among individual) covariances for each trait, and $\mathbf{G}$ and $\mathbf{R}$ are the $p \times p$ matrices modeling genetic and residual covariances among traits within each individual.

We wish to estimate the covariance matrices $\mathbf{G}$ and $\mathbf{R}$. To do so, we assume that any covariance among the observed traits is caused by a number of latent factors. Specifically, we model $k$ latent traits that each linearly relate to one or more of the observed traits. We specify $\mathbf{U}$ and $\mathbf{E}$ via the following hierarchical factor model:

\begin{align}
\label{random_effect_factor_models}
\begin{split}
\mathbf{U} = \mathbf{F}_a \mathbf{\Lambda}^T + \mathbf{\Delta}, &\quad
\mathbf{E} = \mathbf{F}_e \mathbf{\Lambda}^T+ \mathbf{\Xi}\\
\mathbf{F}_a \sim \mbox{MN}_{r,k}(\mathbf{0};\mathbf{A},\mathbf{\Sigma}_a), &\quad
\mathbf{F}_e  \sim \mbox{MN}_{n,k}(\mathbf{0};\mathbf{I}_n,\mathbf{\Sigma}_e) \\
\mathbf{\Delta} \sim \mbox{MN}_{r,p}(\mathbf{0};\mathbf{A},\mathbf{\Psi}_a), & \quad 
\mathbf{\Xi} \sim
\mbox{MN}_{n,p}(\mathbf{0};\mathbf{I}_n,\mathbf{\Psi}_e)\\
\mathbf{\Lambda} &\sim \pi(\theta), \\
\end{split}
\end{align}
\noindent where $\mathbf{\Lambda}$ is a $p \times k$ matrix with each column characterizing the relationship between one latent trait and all observed traits. Just as $\mathbf{U}$ and $\mathbf{E}$ partition the among-individual variation in the \emph{observed} traits into additive genetic effects and residuals in~\eqref{multi_trait_mixedModel}, the matrices $\mathbf{F}_a$ and $\mathbf{F}_e$ partition the among-individual variation in the \emph{latent} traits into additive genetic effects and residuals. $\mathbf{\Sigma}_a$ and $\mathbf{\Sigma}_e$ model the within-individual covariances of $\mathbf{F}_a$ and $\mathbf{F}_e$, which we assume to be diagonal ($\mathbf{\Sigma}_a = \mbox{Diag}(\sigma^2_{a_j}),\mathbf{\Sigma}_e = \mbox{Diag}(\sigma^2_{e_j}))$. $\mathbf{\Psi}_a$ and $\mathbf{\Psi}_e$ are the idiosyncratic (trait-specific) variances of the factor model and are assumed to be diagonal.

In model~\eqref{random_effect_factor_models}, as in any factor model (e.g., \citeNP{West03bayesianfactor}), $\pmb \Lambda$ is not identifiable without adding extra constraints. In general, the factors in $\pmb \Lambda$ can be rotated arbitrarily. This is not an issue for estimating $\mathbf{G}$ itself, but prevents biological interpretations of $\pmb \Lambda$ and makes assessing MCMC convergence difficult. To solve this problem, we introduce constraints on the orientation of $\pmb \Lambda$ though our prior distribution $\pi(\theta)$ specified below, where $\theta$ is a set of hyperparameters. However, even after fixing a rotation, the relative scaling of corresponding columns of $\mathbf{F}_a$, $\mathbf{F}_e$ and $\pmb \Lambda$ are still not well defined. For example, if the $j$th column of $\mathbf{F}_a$ and $\mathbf{F}_e$ are both multiplied by a constant $c$, the same model is recovered if the $j$th column of $\pmb \Lambda$ is multiplied by $1/c$. To fix $c$, we require the column variances $\sigma^2_{a_j}$ and $\sigma^2_{e_j}$ to sum to one, i.e. $\mathbf{\Sigma}_a+ \mathbf{\Sigma}_e = \mathbf{I}_{k}$. Therefore, the single matrix $\mathbf{\Sigma}_{h^2} = \mathbf{\Sigma}_a = \mathbf{I}_{k} - \mathbf{\Sigma}_e$ is sufficient to specify both variances. The diagonal elements of this matrix specify the narrow-sense heritability $(h^2_j = \frac{\sigma^2_{a_j}}{\sigma^2_{a_j} + \sigma^2_{e_j}} = \sigma^2_{a_j})$ of latent trait $j$.

Given the properties of the matrix normal distribution \cite{Dawid:1981gb} and models~\eqref{random_effect_distributions} and~\eqref{random_effect_factor_models} we can recover $\mathbf{G}$ and $\mathbf{R}$ as:

\begin{align}
\label{factors_G_R}
\begin{split}
\mathbf{G} &= \mathbf{\Lambda} \mathbf{\Sigma}_{h^2} \mathbf{\Lambda}^T + \mathbf{\Psi}_a, \\
\mathbf{R} &= \mathbf{\Lambda}(\mathbf{I}_{k}-\mathbf{\Sigma}_{h^2}) \mathbf{\Lambda}^T + \mathbf{\Psi}_e. 
\end{split}
\end{align}
Therefore, the total phenotypic covariance $\mathbf{P} = \mathbf{G} + \mathbf{R}$ is modeled as:

\begin{align}
\label{factors_P}
\mathbf{P} = \mathbf{\Lambda} \mathbf{\Lambda}^T + \mathbf{\Psi}_a+ \mathbf{\Psi}_e.
\end{align}
 
Our specification of the Bayesian genetic sparse factor model in~\eqref{random_effect_factor_models} differs from earlier methods such as the Bayesian genetic factor model of \citeN{deLosCampos:2007ib} in two key respects:

First, in classic factor models, the total number of latent traits is assumed to be small ($k \ll p$). Therefore, equation~\eqref{factors_G_R} would model $\mathbf{G}$ with only $pk + k + p$ parameters instead of $p(p+1)/2$. However, choosing $k$ is a very difficult, unsolved problem, and inappropriate choices can result in highly biased estimates of $\mathbf{G}$ and $\mathbf{R}$ (e.g, \citeNP{Meyer:2008di}). In our model we allow many latent traits but assume that the majority of them are relatively unimportant. This subtle difference is important because it removes the need to accurately choose $k$, instead emphasizing the estimation of the \emph{magnitude} of each latent trait. This model is based on the work by \citeN{Bhattacharya:2011gh}, which they term an ``infinite" factor model. In our prior distribution on the factor loadings matrix $\mathbf{\Lambda}$ (see section Priors), we order the latent traits (columns of $\mathbf{\Lambda}$) in terms of decreasing influence on the total phenotypic variation, and assume that the variation explained by these latent traits decreases rapidly. Therefore, rather than attempt to identify the correct $k$ we instead model the decline in the influence of successive latent traits. As in other factor models, to save computational effort we can truncate $\mathbf{\Lambda}$ to include only its first $k^*<k$ columns because we require the variance explained by each later column to approach zero. The truncation point $k^*$ can be estimated jointly while fitting the model and is flexible (we suggest truncating any columns of $\mathbf{\Lambda}$ defining a module that does not explain $>1\%$ of the phenotypic variation in at least 2 observed traits). Note that $k^*$ conveys little biological information and does not have the same interpretation as $k$ in classic factor models. Since additional factors are expected to explain negligible phenotypic variation, including a few extra columns of $\mathbf{\Lambda}$ to check for more factors is permissible
(e.g., \citeNP{Meyer:2008di}).

Second, we assume that the residual covariance $\mathbf{R}$ has a factor structure and that the same latent traits underly both $\mathbf{G}$ and $\mathbf{R}$. Assuming a constrained space for $\mathbf{R}$ is uncommon in multivariate genetic estimation. For example, \citeN{deLosCampos:2007ib} fit an unconstrained $\mathbf{R}$, although they used an informative inverse Wishart prior \cite{Gelman:2006uw} and only consider five traits. The risk of assuming a constrained $\mathbf{R}$ is that poorly modeled phenotypic covariance ($\mathbf{P}=\mathbf{G}+\mathbf{R}$) can lead to biased estimates of genetic covariance under some circumstances \cite{Jaffrezic:2002tj,Meyer:2008di}. 

However, constraining $\mathbf{R}$ is necessary in high-dimensional settings to prevent the number of modeling parameters from increasing exponentially, and we argue that modeling $\mathbf{R}$ as we have done is biologically justified. Factor models fitting low numbers of latent factors are used in many fields because they accurately model phenotypic covariances. Reasonable constraints on $\mathbf{R}$ have been applied successfully in previous genetic models. One example is in the Direct Estimation of Genetic Principle Components model of \citeN{Kirkpatrick:2004bv}. These authors model only the first $m_E$ eigenvectors of the residual covariance matrix. Our model for $\mathbf{R}$ is closely related to models used in random regression analysis of function-valued traits (e.g., \citeNP{Kirkpatrick:1989wx,Pletcher:1999uy,Jaffrezic:2002tj,Meyer:2005dl}). In those models, $\mathbf{R}$ is modeled as a permanent environmental effect function plus independent error. The permanent environmental effect function is given a functional form similar to (or more complex than) the genetic function. In equation~\eqref{random_effect_factor_models}, $\mathbf{F}_e$ is analogous to this permanent environmental effect (but across different traits rather than the same trait measured through time), with its functional form described by $\mathbf{\Lambda}$, and $\mathbf{\Xi}$ is independent error. Since both $\mathbf{F}_a$ and $\mathbf{F}_e$ relate to the observed phenotypes through $\mathbf{\Lambda}$, the functional form of the residuals ($\mathbf{e}_i$) in our model is at least as complex as the genetic functional form (and more complex whenever $h^2_j = 0$ for some factors). 
 
The biological justification of our approach is that the factors represent latent traits, and just like any other trait their value can partially be determined by genetic variation. For example, the activity of developmental pathways can have a genetic basis and can also be determined by the environment. The latent traits determine the phenotypic covariance of the measured traits, and their heritability determines the genetic covariance. In genetic experiments, some of these latent traits (e.g., measurement biases) might be variable, but not have a genetic component. We expect that some factors will contribute to $\mathbf{R}$ but not $\mathbf{G}$, so $\mathbf{R}$ will have a more complex form \cite{Meyer:2008di}. 

We examine the impact of our prior on $\mathbf{R}$ through simulations below. When our assumptions regarding $\mathbf{R}$ do not hold, the prior will likely lead to biased estimates. For example, measurement biases might be low-dimensional but not sparse. However, we expect that for many general high-dimensional biological datasets this model will be useful and can provide novel insights. In particular, by directly modeling the heritability of the latent traits, we can predict their evolution.

 \subsection{Priors:}
\label{prior}
Modeling high-dimensional data requires some prior specification or penalty/regularization for accurate and stable parameter estimation  \cite{elements,West03bayesianfactor,Poggio03themathematics}. For our model this means that constraints on $\mathbf{G}$ and $\mathbf{R}$ are required. We impose constraints through highly informative priors on $\pmb \Lambda$. Our priors are motivated by the biological assumptions that variation in underlying developmental processes such as gene networks or metabolic pathways give rise to to genetic and residual covariances. 

This implies:\\
\noindent (1) The biological system has limited complexity: a small number of latent traits (e.g., developmental pathways) or measurement biases are relevant for trait variation. For the model this means that the number of factors retained in $\pmb \Lambda_{k^*}$ is low ($k^* \ll p$). \\
\noindent (2) Each underlying latent trait affects a limited number of the observed traits. For the model this means the factor loadings (columns of $\pmb \Lambda$) are sparse (mostly near zero).

We formalize the above assumptions by priors on $ \pmb{\Lambda}$ that impose sparsity (formally, shrinkage towards zero) and few highly influential latent traits \cite{Bhattacharya:2011gh}. This prior is specified as a  hierarchical distribution on each element $\lambda_{ij}$ of $\pmb \Lambda$:

\begin{align}
\label{prior_Lambda}
\begin{split}
\lambda_{ij} &\mid \phi_{ij}, \tau_j \sim \mbox{N} \left( 0,\phi_{ij}^{-1}\tau_j^{-1} \right),\; i=1\dots  p,\;j = 1\dots  k\\
\phi_{ij} &\sim \mbox{Ga}(\nu/2,\nu/2), \\
\tau_j &= \prod \limits_{l=1}^m  \delta_l, \quad 
\delta_1 \sim \mbox{Ga}(a_1,b_1), \quad
\delta_l \sim \mbox{Ga}(a_2,b_2) \mbox{ for } l = 2\dots  k.
\end{split}
\end{align}

\noindent The hierarchical prior is composed of three levels:
\noindent (a) We model each $\lambda_{ij}$ (which specifies how trait $i$ is related to latent trait $j$) with a normal distribution. Based on assumption (2), we expect most $\lambda_{ij} \approx 0$. A normal distribution with a fixed variance parameter is not sufficient to impose this constraint.
\noindent (b) We model the the precision (inverse of the variance) of each loading element $\lambda_{ij}$ with
the parameter $\phi_{ij}$ drawn from a gamma distribution. This normal-gamma mixture distribution (conditional on 
$\tau_j$)  is commonly used to impose sparsity \cite{Tipping2001,Neal1996} as the marginal distribution 
on $\lambda_{ij}$ takes the form of a Student's-\textit{t} distribution with $\nu$ degrees of freedom and is 
heavy-tailed. The loadings are concentrated near zero, but occasional large magnitude values are permitted. This prior specification is conceptually similar to the widely-used Bayesian Lasso (Park and Casella 2008). 
\noindent (c) The parameter $\tau_j$ controls the overall variance explained by factor $j$ (given by: $\pmb \lambda_j^T \pmb \lambda_j$ where $\pmb \lambda_j$ is the $j$th column of $\mathbf{\Lambda}$) by shrinking the variance towards zero as $m \rightarrow \infty$. The decay in the variance is enforced by increasing the precision on
the normal distribution of each $\lambda_{ij}$ as $j$  increases so $|\lambda_{ij}| \rightarrow 0$. The sequence $\{\tau_j\}$ is formed from the cumulative product of $\{\delta_1\dots \delta_k\}$ each modeled with a gamma distribution, and will be stochastically increasing as long as $a_2 > b_2$. This means that the variance of $\lambda_{ij}$ will stochastically decrease and higher-indexed columns of $\pmb \Lambda$ will be less likely to have any large magnitude elements. This decay ensures that it will be safe to truncate $\pmb \Lambda$ at some sufficiently large $k^*$ because columns $k > k^*$ will (necessarily) explain less variance. 	

	The prior distribution on $\{\tau_j\}$ (and therefore $\{\delta_1\dots \delta_k\}$) is a key modeling decision as this parameter controls how much of the total phenotypic variance we expect each successive factor to explain. Based on assumption (1), we expect that few factors will be sufficient to explain total phenotypic variation, and thus $\{\tau_j\}$ will increase rapidly. However, relatively flat priors on $\{\delta_2\dots \delta_k\}$ (e.g., $a_2 = 3, b_2=1$), which allow some consecutive factors to be of nearly equal magnitude, appear to work well in simulations. 
	
A discrete set of values in the unit interval were specified as the prior for the heritability of each of the common factor traits. This specification was selected for computational efficiency and to give $h^2_j=0$ positive weight in the prior. We find the following discrete distribution works well:

\begin{align}
\label{prior_h2}
\pi(h^2_j = 0) = 0.5,\quad 
\pi(h^2_j = l / n_h) = 1/(2(n_h-1)), \mbox{ where } l = 1 \dots (n_h-1)
\end{align}
\noindent where $n_h$ is the number of points to evaluate $h^2_j$. In analyses reported here, we set $n_h = 100$. This prior gives equal weight to $h^2_j = 0$ and $h^2_j > 0$ because we expect several factors (in particular, those reflecting measurement error) to have no genetic variance. In principle, we could place a continuous prior on the interval $[0,1]$, but no such prior would be conjugate, and developing a MCMC sampler would be more difficult.

We place inverse gamma priors on each element of the diagonals of the genetic and residual idiosyncratic variances: $\mathbf{\Psi}_a$ and $\mathbf{\Psi}_e$. Priors on each element of $\pmb \beta$ are normal distributions with very large ($>10^6$ ) variances.

\subsection{Implementation:}

Inference in the above model uses an adaptive Gibbs sampler for which
we provide detailed steps in the appendix. The code has been
implemented in Matlab$\textsuperscript{\textregistered}$ and can be
found at the website (http://stat.duke.edu/$\sim$sayan/quantmod.html).

\subsection{Simulations:}
We present a simulation study of high-dimensional traits measured on the offspring of a balanced paternal half-sib breeding design. We examined ten scenarios (Table~\ref{table:simulation_setup}), each corresponding to different models for the matrices $\mathbf{G}$ and $\mathbf{R}$ to evaluate the impact of the modeling assumptions specified by our prior. For each scenario we simulated parameters and trait values of individuals from model~\eqref{multi_trait_mixedModel} with $\mathbf{Z} = \mathbf{I}_n$, $\mathbf{B} = \mathbf{0}_p$, and $\mathbf{X}$ a single column of ones representing the trait means. 

	Scenarios \emph{a}-\emph{c} were designed to test the ability of the model to accurately estimate $\mathbf{G}$ and $\mathbf{P}$ given 10, 25 or 50 important factors, respectively, for 100 traits. Latent factor traits $1\dots 5$, $1\dots 15$, or $1 \dots 30$, respectively, were assigned a heritability ($h^2_j$) of 0.5 and contributed to both $\mathbf{G}$ and $\mathbf{R}$. The remaining factors ($6 \dots 10$, $16 \dots 25$, or $31 \dots 50$, respectively) were assigned a heritability of 0.0 and only contributed to $\mathbf{R}$. To make the covariance matrices biologically reasonable, we chose each factor to be sparse: only 3-25 of the 100 traits were allowed to ``load" on each factor. These loadings were drawn from standard normal distributions. The idiosyncratic variances $\pmb \Psi_a$ and $\pmb \Psi_e$ were set to $0.2\times\mathbf{I}_p$. Therefore, trait-specific heritabilties ranged from 0.0-0.5, with the majority towards the upper limit. Each simulation included 10 offspring from 100 unrelated sires.
	
	Scenarios \emph{d}-\emph{e} were designed to test the effects of deviations of $\mathbf{R}$ from the modeling assumptions, since it is known that inappropriately modeled residual variances can lead to biased estimates of $\mathbf{G}$ (e.g., \citeNP{Jaffrezic:2002tj,Meyer:2007uz}). Scenarios were identical to \emph{a} except the $\mathbf{R}$ matrix did not have a sparse factor form. In scenario \emph{d}, $\mathbf{R}$ was assumed to follow a factor structure with 10 factors, but five of these factors (numbers $6 \dots 10$, i.e., those with $h^2_j = 0.0$) were not sparse (i.e., all factor loadings were non-zero). This might occur, for example, if the non-genetic factors in the residual were caused by measurement error. In scenario \emph{e}, $\mathbf{R}$ did not follow a factor structure at all, but was drawn from a central Wishart distribution with $p+1$ degrees of freedom. 
	
	Scenarios \emph{f}-\emph{g} were designed to evaluate the performance of the model for different numbers of traits. These scenarios were identical to scenario \emph{a} except 20 or 1,000 (scenarios \emph{f} and \emph{g}, respectively) traits were simulated. As in scenario \emph{a}, all factors were sparse: In scenario \emph{f}, each simulated factor had non-zero loadings for 3-5 traits. In scenario \emph{g}, each simulated factor had non-zero loadings for 30-250 traits.  
	
	Scenarios \emph{h}-\emph{j} were designed to evaluate the performance of the model for experiments of different size, and also to test different latent factor heritabilities. Simulations were generated as in scenario \emph{a}, except that the five genetic factors were assigned heritabilites of 0.9, 0.7, 0.5, 0.3 and 0.1, the number of sires was set to 50, 100 or 500, and the number of offspring per sire was set to 5 (for simulation \emph{h} only).
	
	To fit the simulated data, we set the prior hyperparameters in the model to: 
$\nu=3, a_1=2, b_1 = 1/20, a_2 = 3, b_2 = 1$. We ran the Gibbs sampler for 12,000 iterations, discarded the first 10,000 samples as burn-in, and collected 1,000 posterior samples with a thinning rate of two.

\ctable[
	cap     = {Simulation parameters},
	caption = {Simulation parameters. Eight simulations were designed to demonstrate the capabilities of the Bayesian genetic sparse factor model. Scenarios \emph{a}-\emph{c} test genetic and residual covariance matrices composed of different numbers of factors. Scenarios \emph{d}-\emph{e} test residual covariance matrices that are not sparse. Scenarios \emph{f}-\emph{g} test different numbers of traits. Scenarios \emph{h}-\emph{j} test different sample sizes. All simulations followed a paternal half-sib breeding design. Each simulation was run 10 times.},
	label   = {table:simulation_setup},
]{lccc  cc  cc  ccc}{
	\tnote[a]{Sparse factor model for \textbf{R}. Each simulated factor loading ($\lambda_{I'm}$) had a $3\%-25\%$ chance of not equaling zero.}
	\tnote[b]{Factor model for \textbf{R}. Residual factors (those with $h^2_j = 0$) were not sparse ($\lambda_{ij} \neq 0 \; \forall \; i \in 1\dots p$).}
	\tnote[c]{\textbf{R} was simulated from a Wishart distribution with $p+1$ degrees of freedom and inverse scale matrix $\frac{1}{p}\mathbf{I}_p$. All factors were assigned a heritability of 1.0}
	\tnote[d]{In each column, factors are divided between those with positive and zero heritability. The number in parentheses provides the number of factors with the given heritability.}
}{ \FL
&\multicolumn{3}{c}{\# factors} & \multicolumn{2}{c}{\textbf{R} type} & \multicolumn{2}{c}{\# traits} & \multicolumn{3}{c}{Sample size} \NN
 & \emph{a} & \emph{b}& \emph{c} & \emph{d} & \emph{e} & \emph{f} & \emph{g} & \emph{h} &\emph{i}&\emph{j}\NN \cmidrule(r){2-4} \cmidrule(r){5-6} \cmidrule(r){7-8}\cmidrule(r){9-11}  
\textbf{G} and \textbf{R} & & & & & & &  &&&\NN  \cmidrule(r){1-1} 
\quad \# traits & 100 & 100 & 100 & 100 & 100 & 20 & 1000 & 100 & 100 & 100  \NN
\quad Residual type & SF\tmark[a] & SF & SF & F\tmark[b] & Wishart\tmark[c] & SF & SF & SF & SF & SF \NN
\quad \# factors & 10 & 25 & 50 & 10 & 5 & 10 & 10& 10 & 10 & 10\NN
\quad \multirow{2}{*}{$h^2$ of factors\tmark[d] } & 0.5(5) & 0.5(15) & 0.5(30) & 0.5(5) & 1.0(5) & \multicolumn{2}{c}{0.5(5)} & \multicolumn{3}{c}{0.9-0.1(5)} \NN 
\quad & 0.0(5) & 0.0(10) & 0.0(20) & 0.0(5) & & \multicolumn{2}{c}{0.0(5)} &\multicolumn{3}{c}{0.0(5)} \NN \NN
Sample Size & & & & & & && &&  \NN  \cmidrule(r){1-1} 
\quad \# sires & 100 & 100 & 100 & 100 & 100 & 100 & 100 & 50 & 100 & 500\NN
\quad \# offspring/sire & 10 & 10 & 10 & 10 & 10 & 10 & 10 & 5 & 10 & 10 \LL}

\subsubsection{Evaluation}
We calculated a number of statistics from each simulation to quantify the error in the model fits produced by our Bayesian genetic sparse factor model. For each statistic, we compared the posterior mean of a model parameter to the true value specified in the simulation. 

	First, as a sanity check, we compared the accuracy of our method to a methods of moments estimate of $\mathbf{G}$ calculated as $\mathbf{G}_m = 4(\mathbf{B}-\mathbf{W})/n$ where $\mathbf{B}$ and $\mathbf{W}$ are the between and within sire matrices of mean squares and cross products and $n$ is the number of offspring per sire. We compared the accuracy of the moments estimator $\mathbf{G}_m$ to the posterior mean $\hat{\mathbf{G}}$ from our model by calculating the Frobenius norm of the differences $\mathbf{G}_m - \mathbf{G}$ and $\hat{\mathbf{G}}-\mathbf{G}$. 

	The Frobenius norm measure above simply quantifies the total (sum of square) error in each pairwise covariance estimate. However, the geometry of $\mathbf{G}$ is more important for predicting evolution \cite{Walsh:2009gg}. We evaluated the accuracy of the estimated $\textbf{G}$ matrix by comparing the $k$-dimensional subspace of $\mathbb{R}^p$ with the majority of the variation in $\mathbf{G}$ to the corresponding subspace for the posterior mean estimate $\hat{\mathbf{G}}$. For this, we calculated the Krzanowski subspace comparison statistic \cite{Krzanowski:1979cx,Blows:2004ui}, which is the sum of the eigenvalues of the matrix $\mathbf{S} = \hat{\mathbf{G_k}}^T\mathbf{G_k}\mathbf{G_k}^T\hat{\mathbf{G_k}}$, where $\hat{\mathbf{G_k}}$ is the subspace spanned by the eigenvectors with the $k$ largest eigenvalues of the posterior mean of $\mathbf{G}$, and $\mathbf{G_k}$ is the corresponding subspace of the true (simulated) matrix. This statistic will be zero for orthogonal (non-overlapping) subspaces, and will equal $k$ for identical subspaces. The accuracy of the estimated $\mathbf{P}$ was calculated similarly. For each comparison, $k$ was chosen as the number of factors used in the construction of the simulated matrix (Table~\ref{table:simulation_setup}), except in scenario E with the Wishart-distributed $\mathbf{R}$ matrix. Here, we set the $k$ for $\mathbf{P}$ at 19 which was sufficient to capture $>99\%$ of the variation in most simulated $\mathbf{P}$ matrices. 

	We evaluated the accuracy of latent factors estimates in two ways. First, we calculated the magnitude of each factor as $|\pmb{\lambda}_j|^2$ where $|\cdot|$ is the L$_2$-norm. This quantifies the phenotypic variance across all traits explained by each factor. We then counted the number of factors that explained $>0.1\%$ of total phenotypic variance. Such factors were termed ``large factors". Second, we matched each of the simulated factors ($\pmb \lambda_j$) to the most similar estimated factor ($\pmb \lambda_{j^*}$) and calculated the estimation error in each simulated factor as the angle between the two vectors. Smaller angles correspond to more accurately identified factors.
\subsection{Gene expression analysis:}
	We downloaded gene expression profiles and measures of competitive fitness of 40 wild-derived lines of \textit{Drosophila melanogaster} from ArrayExpress (accession: E-MEXP-1594) and the Drosophila Genetic Reference Panel (DGRP) website (http://dgrp.gnets.ncsu.edu/) \citeNP{Ayroles:2009gd}). A line's competitive fitness \cite{GRKnight:1957ww,Hartl:1979ub} measures the percentage of offspring bearing the line's genotype recovered from vials seeded with a known proportion of adults from that line and adults of a reference line. We used our Bayesian genetic sparse factor model to infer a set of latent factor traits underlying the among-line gene expression covariance matrix for a subset of the genes and the among line covariance between each gene and competitive fitness. These latent factors are useful because they provide insight into what genes and developmental or molecular pathways underlie variation in competitive fitness.
	
	We first normalized the processed gene expression data to correspond to the the analyses of the earlier paper and then selected the 414 genes that \citeN{Ayroles:2009gd} identified as having a plausible among-line covariance in competitive fitness. In this dataset, two biological replicates of male and female fly collections from each line were analyzed for whole-animal RNA expression. The competitive fitness measurements were the means of 20 competitive trials done with sets of flies from these same lines, but not the same flies used in the gene expression analysis. Gene expression values for the samples measured for competitive fitness and competitive fitness values for the samples measured for gene expression were treated as missing data (see Appendix). We used our model to estimate the G-matrix of the genes (the covariance of line effects). Following the analyses of \citeN{Ayroles:2009gd}, we included a fixed effect of sex, and independent random effects of the sex:line interaction for each gene. No sex or sex:line effects were fit for competitive fitness itself as this value is measured at the level of the line, not individual flies.

We set the prior hyperparameters as above, and ran our Gibbs sampler for 40,000 iterations, discarded the first 20,000 samples as a burn-in period, and collected 1,000 posterior samples of all parameters with a thinning rate of 20. 

\section{Results}	 
\subsection{Simulation example:} 

The Bayesian genetic sparse factor model's estimates of genetic covariances across the 100 genes were considerably more accurate than estimates based on unbiased methods of moments estimators. In scenario \emph{a}, for example, the mean Frobenius norm was 13.9 for the moments estimator and 6.3 for the Bayesian genetic sparse factor model's posterior mean, a $54\%$ improvement. 

\begin{figure}[htbp!]
\begin{center}
\includegraphics{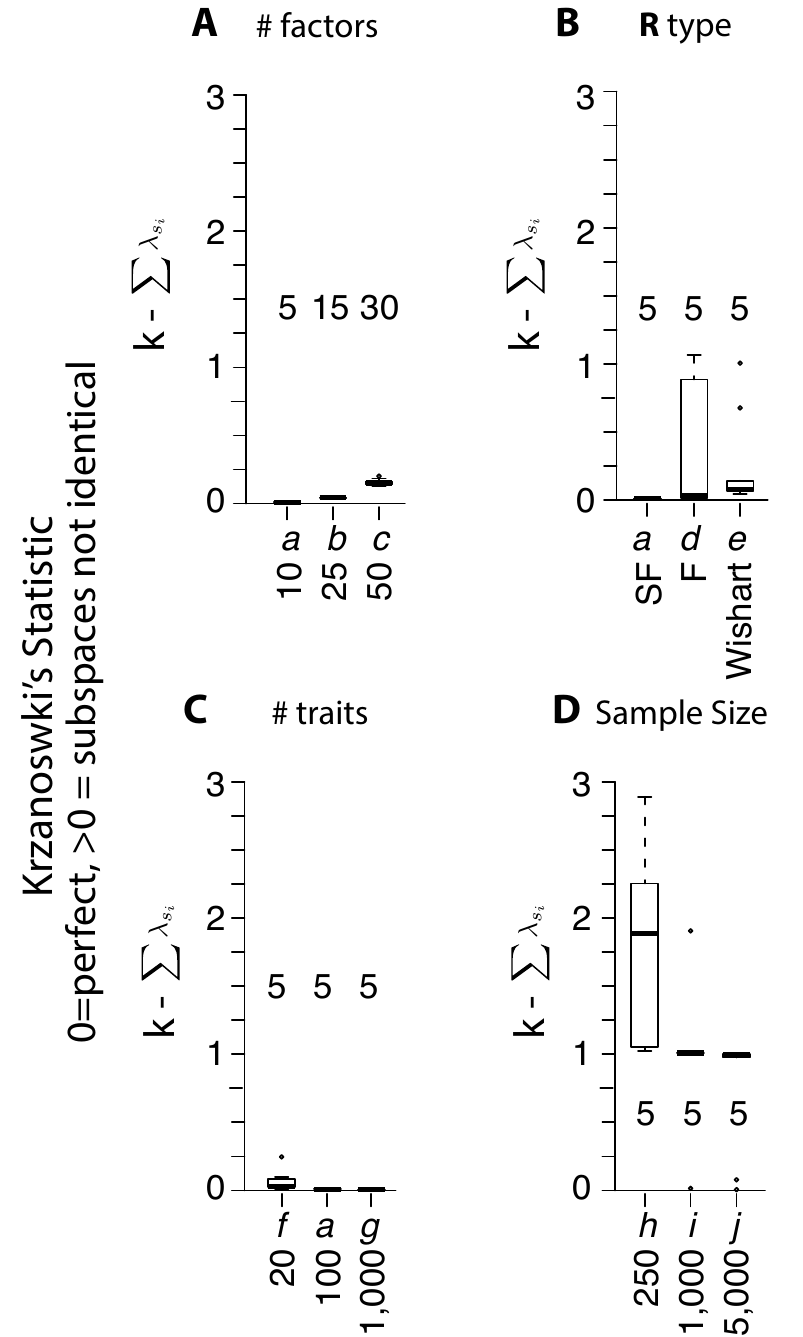}
\caption[Krzanowski substance comparison statistics for G by scenario]{ \textbf{The Bayesian genetic sparse factor model accurately estimates the dominant subspace of high-dimensional G matrices.} Each subplot shows the distribution of Krzanowski's statistics ($\sum \lambda_{s_i}$, \citeNP{Krzanowski:1979cx,Blows:2004ui}) calculated for posterior mean estimates of $\mathbf{G}$ across a related set of scenarios. Plotted values are $k - \sum \lambda_{s_i}$ so that statistics are comparable across scenarios with different subspace dimensions. On this scale, identical subspaces have a value of zero and values increase as the subspaces diverge. The value of $k$ used in each scenario is listed inside each boxplot. The difference from zero roughly corresponds to the number of eigenvectors of the true subspace missing from the estimated subspace. Different parameters were varied in each set of simulations as listed below each box. \textbf{A}. Increasing numbers of simulated factors. \textbf{B}. Different properties of the $\mathbf{R}$ matrix. ``SF": a sparse-factor form for $\mathbf{R}$, ``F": a (non-sparse) factor form for $\mathbf{R}$, ``Wishart": $\mathbf{R}$ was sampled from a Wishart distribution. \textbf{C}. Different numbers of traits. \textbf{D}. Different numbers of sampled individuals. Note that in these scenarios, factor $h^2$s ranged from 0.0 to 0.9. Complete parameter sets describing each simulation are described in Table~\ref{table:simulation_setup}.}
\label{fig:Krzanowski_G}
\end{center}
\end{figure}

\begin{figure}[htbp!]
\begin{center}
\includegraphics{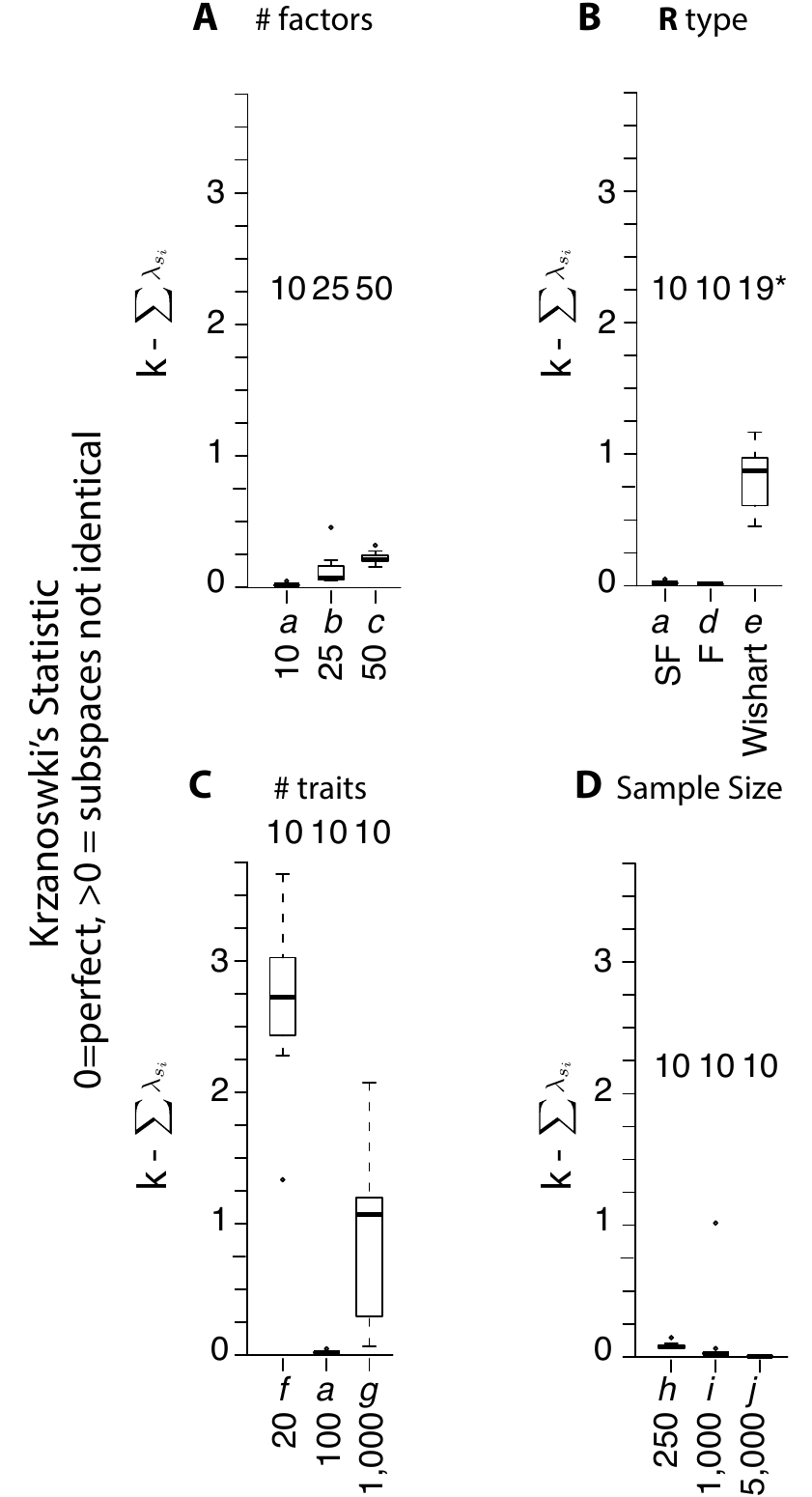}
\caption[Krzanowski substance comparison statistics for P by scenario]{ \textbf{$\mathbf{P}$-matrix subspaces were accurately recovered.} This figure is identical to Figure 1 but for $\mathbf{P}$. Each subplot shows the distribution of Krzanowski's statistics ($\sum \lambda_{s_i}$) calculated for posterior mean estimates of $\mathbf{P}$ across a related set of scenarios. The value of $k$ used in each scenario is listed inside each boxplot. The parameter varied in each set of simulations is described at the bottom. (A) Increasing numbers of simulated factors. (B) Different properties of the $\mathbf{R}$ matrix. ``SF": a sparse-factor form for $\mathbf{R}$, ``F": a (non-sparse) factor form for $\mathbf{R}$, ``Wishart": $\mathbf{R}$ was sampled from a Wishart distribution. In scenario \emph{e}, the residual matrix did not have a factor form. Therefore, we chose $k=19$ for the phenotypic covariance matrix because the corresponding eigenvectors each explained $>1\%$ of total phenotypic variation. (C) Different numbers of traits. (D) Different numbers of sampled individuals. Complete parameter sets describing each simulation are described in Table 1.}
\label{fig:Krzanowski_P}
\end{center}
\end{figure}

Our model also produced accurate estimates of the subspaces containing the majority of variation in both $\mathbf{G}$ and $ \mathbf{P}$. Figure~\ref{fig:Krzanowski_G} shows the distribution of Krzanowski's subspace similarity statistics ($\sum \lambda_{s_i}$) for $\mathbf{G}$ in each scenario (Subspace statistics for $\mathbf{P}$ are shown in Figure~\ref{fig:Krzanowski_P}). Krzanowski's statistic roughly corresponds to the number of eigenvectors of the true subspace missing from the estimated subspace. We plot $k-\sum \lambda_{s_i}$ so that the values are comparable across simulations with different $k$. The Krzanowski statistics were all close to $k$, rarely diverging even one unit except in scenarios \emph{h}-\emph{j} where one of the genetic factors was particularly difficult to estimate. This indicates that the subspaces of both matrices were largely recovered across all scenarios. However, Krzanowski's difference (relative to $k$) for $\mathbf{G}$ increased slightly for larger numbers of factors (Figure~\ref{fig:Krzanowski_G}A), if $\mathbf{R}$ did not follow a factor structure (Figure~\ref{fig:Krzanowski_G}B), if few traits were measured (Figure~\ref{fig:Krzanowski_G}C), or if the sample size was small (Figure~\ref{fig:Krzanowski_G}D). Some simulations when the latent factors of $\mathbf{R}$ were not sparse also caused slight subspace errors (scenario \emph{d}, Figure~\ref{fig:Krzanowski_G}B). In scenarios \emph{h}-\emph{j}, the 10th factor was assigned a heritability of only $10\%$ and so the subspace spanned by the first five eigenvectors of estimated $\mathbf{G}$ matrices often did not include this vector. This effect was exacerbated at low sample sizes. Krzanowski's statistics for $\mathbf{P}$ followed a similar pattern  (Figure~\ref{fig:Krzanowski_P}), except that the effect of a lack of a factor structure for $\mathbf{R}$ were more pronounced (Figure~\ref{fig:Krzanowski_P}B), as was the reduced performance for different numbers of traits (Figure~\ref{fig:Krzanowski_P}C). 

\begin{figure}[htbp!]
\begin{center}
\includegraphics{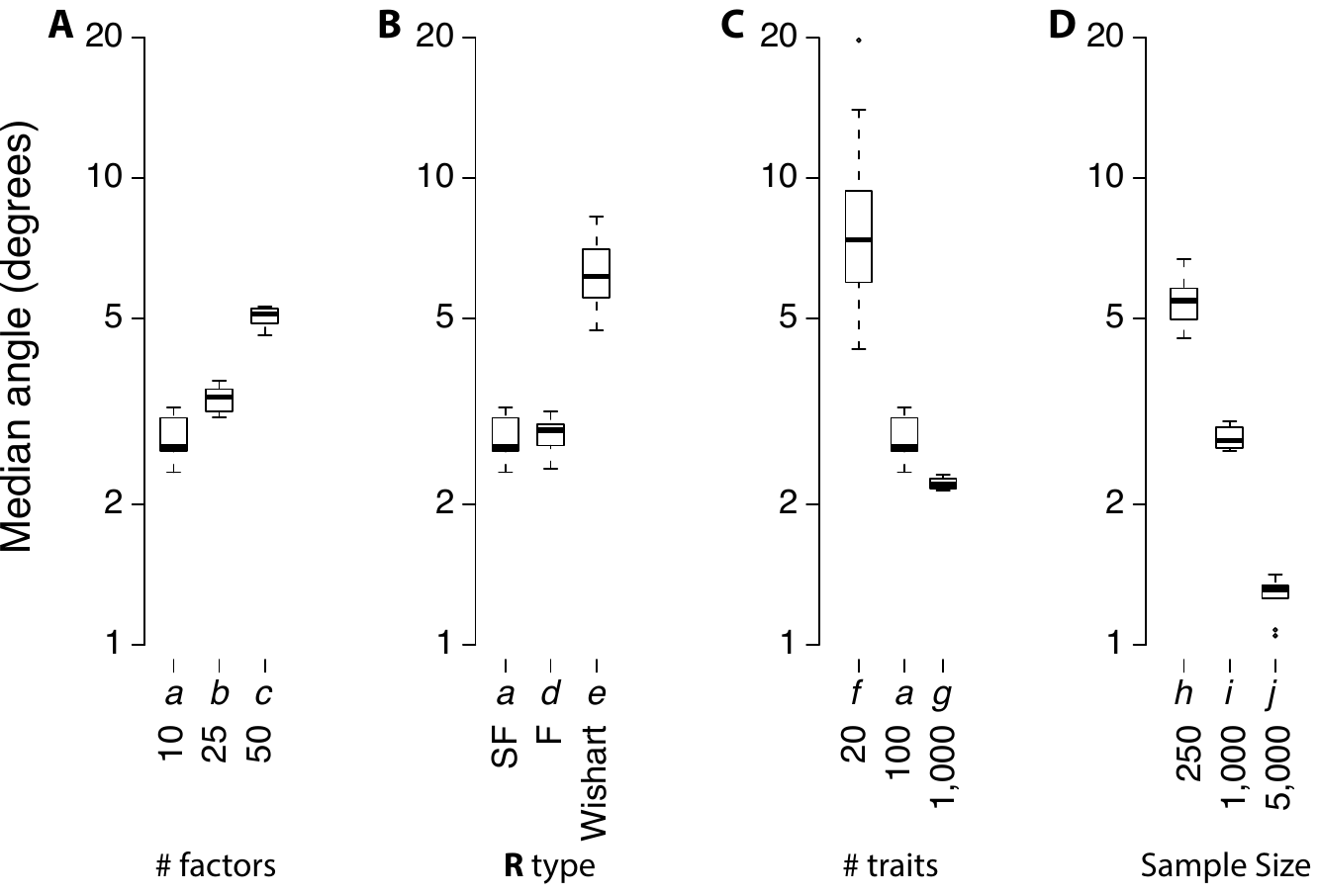}
\caption[Error in latent factor trait loadings by scenario]{ \textbf{Latent factors were accurately recovered in most simulations.} The true factors in each simulation were matched to the most similar estimated factor by calculating the vector angles between each true factor and each estimated factor. The median error angle for each true factor in each simulation is plotted. Boxplots show the distribution of median error angles by scenario. Two identical vectors have an angle of zero. Completely orthogonal vectors have an angle of 90. \textbf{A}. Increasing numbers of simulated factors. \textbf{B}. Different properties of the $\mathbf{R}$ matrix. \textbf{C}. Different numbers of traits. \textbf{D}. Different numbers of sampled individuals.}
\label{fig:factor_angles}
\end{center}
\end{figure}

Even though the number of latent factors is not an explicit model parameter, the number of ``large factors" fit in each scenario was always close to the the true number of simulated factors (except for scenario \emph{e} where $\mathbf{R}$ did not have a factor form). Median numbers of estimated ``large factors" are given in Table~\ref{table:large_factor_counts}. The identities of the factors identified by our model were also accurate. Figure~\ref{fig:factor_angles} shows the distribution of error angles between the true factors and their estimates for each scenario. Median angles were greater for larger numbers of latent factors (Figure~\ref{fig:factor_angles}A), if $\mathbf{R}$ did not follow a factor structure (scenario \emph{e}, Figure~\ref{fig:factor_angles}B), or for smaller sample sizes (small numbers of individuals or small numbers of traits, scenarios \emph{f},\emph{h}, Figure~\ref{fig:factor_angles}C-D). For scenarios \emph{d} and \emph{e}, angles are shown only for the factors that contributed to $\mathbf{G}$ (factors 1-5). The residual factors for these scenarios were not well defined (In scenario \emph{d}, factors 6-10 were not sparse and thus were only identifiable up to an arbitrary rotation by any matrix $\mathbf{H}$ such that $\mathbf{H}\mathbf{H}^T = \mathbf{I}$ \cite{Meyer:2009jm}. In scenario \emph{e}, the residual matrix did not have a factor form).

\ctable[	cap     = {Number of large factors recovered},
	caption = {Number of large factors recovered in each scenario. Each scenario was simulated 10 times. Factor magnitude was calculated as the L$_2$-norm of the factor loadings, divided by the total phenotypic variance across all traits. Factors explaining $>0.1\%$ of total phenotypic variance were considered large.},
	label   = {table:large_factor_counts},
	mincapwidth=6in
]{cc|ccc}{	\tnote[a]{In scenario E, the residual matrix did not have a factor form.}
}{ \FL	
\multicolumn{2}{c}{Scenario} & Expected & Median & Range\ML
\multirow{3}{*}{\# factors} & \emph{a} & 10 & 10 & (10,10)\\ 
& \emph{b} & 25 & 25 & (23,25)\\ 
& \emph{c} & 50 & 49 & (48,50)\ML
\multirow{2}{*}{$\mathbf{R}$ type} & \emph{d} & 10 & 10 & (10,10)\\ 
& \emph{e} & NA\tmark[a] & 56 & (44,66)\ML
\multirow{2}{*}{\# traits} & \emph{f} & 10 & 9 & (8,11)\\ 
& \emph{g} & 10 & 10 & (10,10)\ML
\multirow{3}{*}{Sample size} & \emph{h} & 10 & 10 & (10,10)\\ 
& \emph{i} & 10 & 10 & (10,10)\\ 
& \emph{j} & 10 & 10 & (10,10)\LL}

\begin{figure}[htbp!]
\begin{center}
\includegraphics{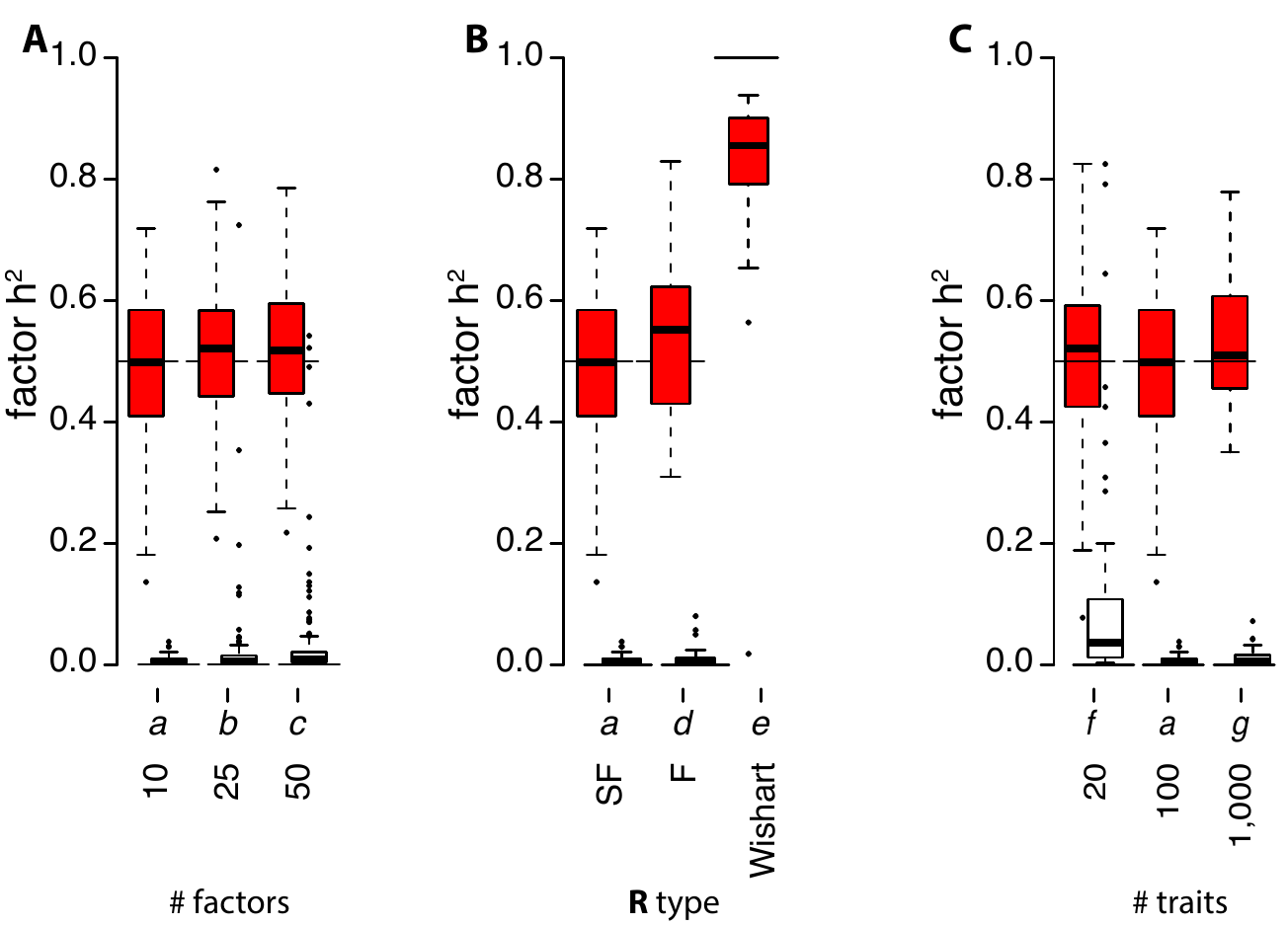}
\caption[Heritability of latent factors]{ \textbf{Latent factor heritabilities were accurately recovered.} Distributions of factor $h^2$ estimates by simulation scenario. Each simulated factor was matched to the estimated factor with most similar trait-loadings as in Figure~\ref{fig:factor_h2s_D}. Thin horizontal lines in each column show the simulated $h^2_j$ values. Each simulated factor was assigned $h^2 = 0.0$ (black) or $0.5$ (red), except in scenario \emph{e} where all five factors were assigned $h^2 = 1$ (red). $h^2$ estimates are are grouped across all 10 simulations of each scenario. (A) Increasing numbers of simulated factors. (B) Different properties of the $\mathbf{R}$ matrix. (C) Different numbers of traits.}
\label{fig:factor_h2s_ABC}
\end{center}
\end{figure}

\begin{figure}[htbp!]
\begin{center}
\includegraphics{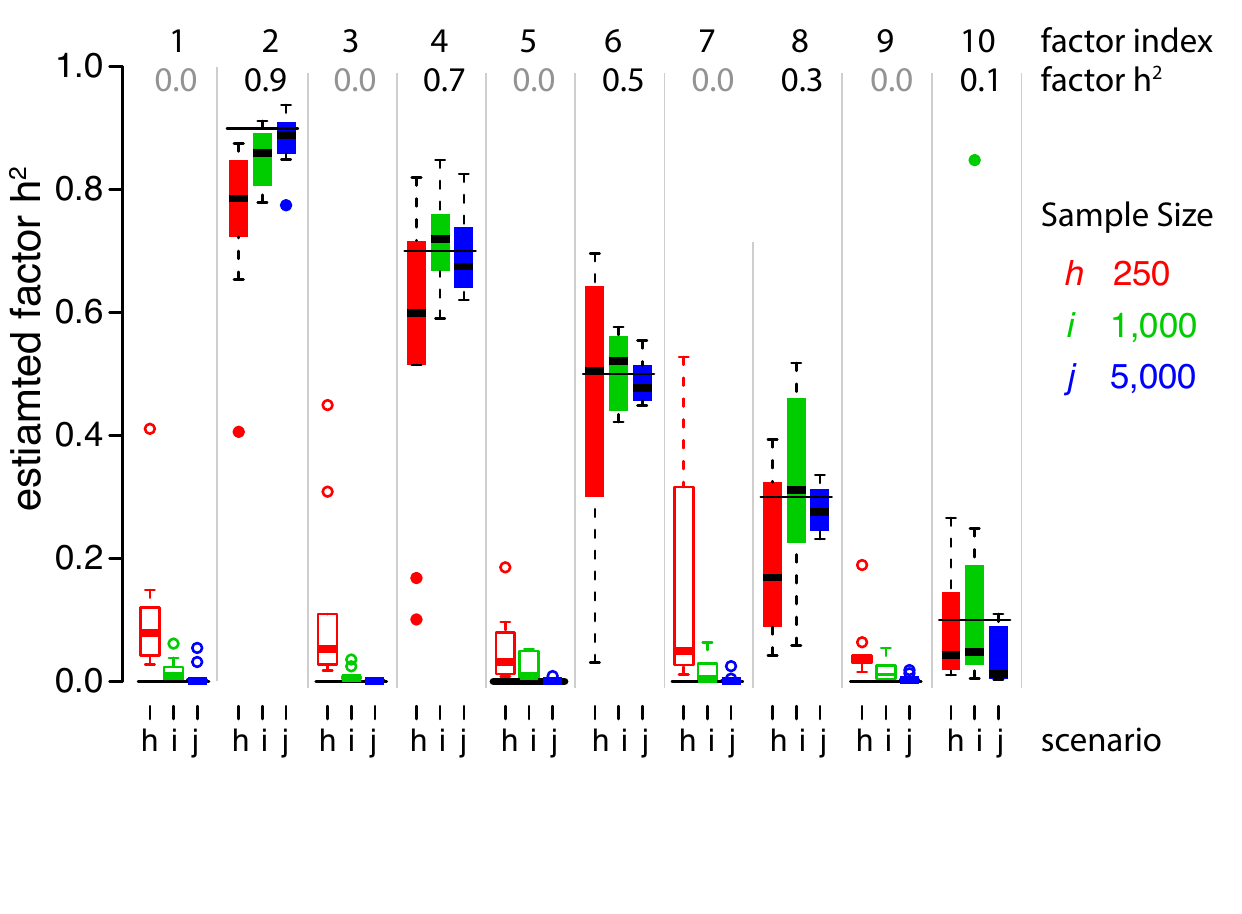}
\caption[Heritability of latent factors]{ \textbf{Latent factor heritabilities were accurately recovered.} Distributions of factor $h^2$ estimates for scenarios \emph{h}-\emph{j}. These scenarios differed in the number of individuals sampled. 10 factors were generated in each simulation and assigned $h^2$s between 0.0 and 0.9. After fitting our factor model to each simulated dataset,  the simulated factors were matched to estimated factors based on the trait-loading vector angles. Each boxplot shows the distribution of $h^2$ estimates for each simulated factor across 10 simulations. Note that the trait-loadings for each factor differed in each simulation; only the $h^2$ values remained the same. Thin horizontal lines in each column show the simulated $h^2_j$ values. Colors correspond to the scenario, and filled boxes/circles are used for factors with $h^2_j > 0$.}
\label{fig:factor_h2s_D}
\end{center}
\end{figure}

\begin{figure}[htbp!]
\begin{center}
\includegraphics{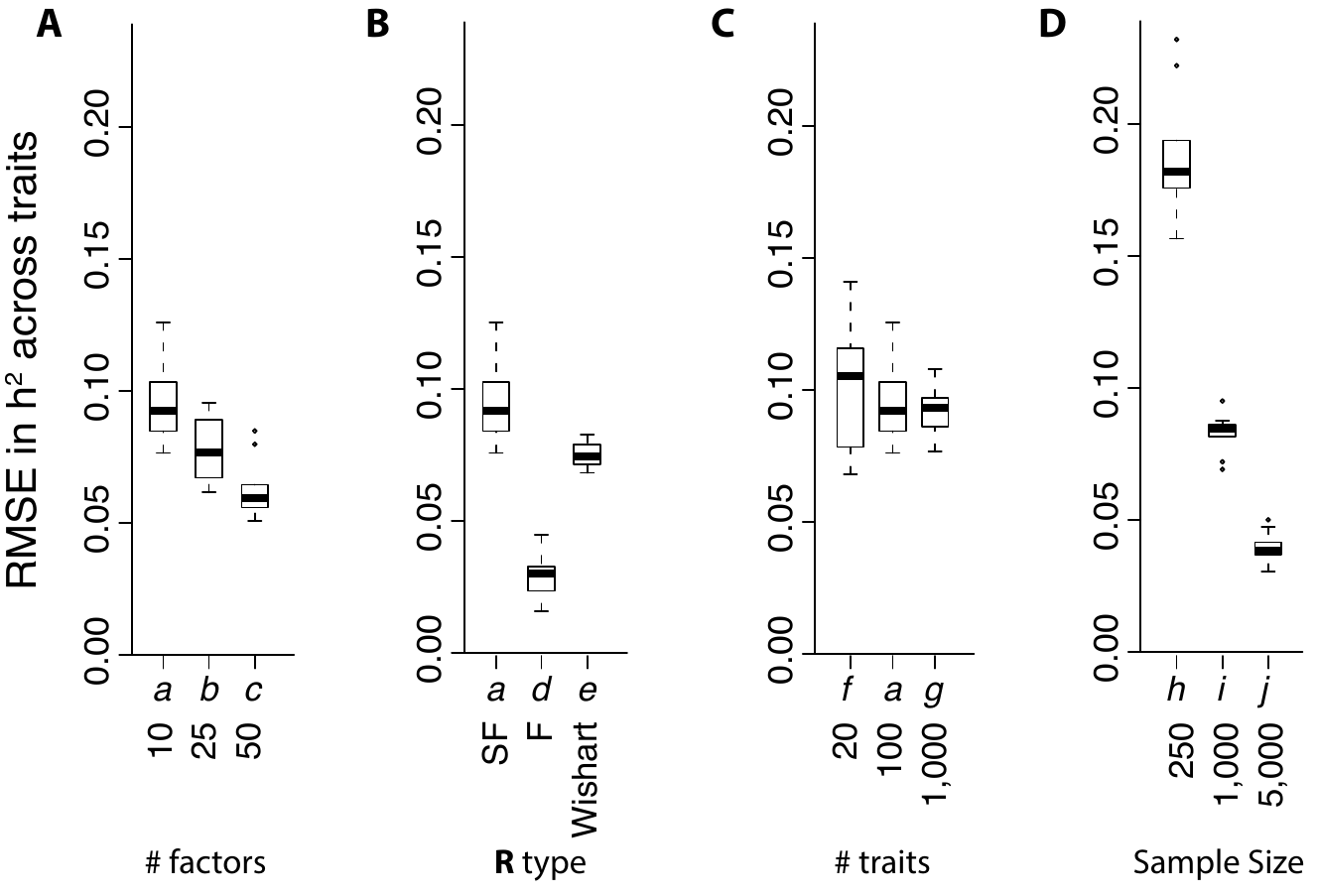}
\caption[Error in trait heritability]{ \textbf{Heritability estimates for each individual trait were accurate.} The heritability of each individual trait was estimated as $h^2_i = \mathbf{G}_{ii}/\mathbf{P}_{ii}$. RMSE $= \sqrt{\frac{1}{p}\sum\limits_{i=1}^p(\hat{h}^2_i - h^2_i)^2}$ was calculated for each simulation. Boxplots show the distribution of RMSE values for each scenario. \textbf{A}. Increasing numbers of simulated factors. \textbf{B}. Different properties of the $\mathbf{R}$ matrix. \textbf{C}. Different numbers of traits. \textbf{D}. Different numbers of sampled individuals. }
\label{fig:trait_h2s}
\end{center}
\end{figure}

	Finally, the genetic architectures of the unmeasured latent traits (factors) and the measured traits were accurately estimated. For scenarios \emph{a}-\emph{d} and \emph{f}-\emph{g}, each latent factor was assigned a heritability of either 0.5 or 0.0. Heritability estimates for factors with simulated heritability of 0.5 were centered around 0.5, and with $>50\%$ between 0.4 and 0.6 (Figure~\ref{fig:factor_h2s_ABC}). There was little difference in performance for these factors across scenarios with different numbers of factors, different residual properties, or different numbers of traits (Figure~\ref{fig:factor_h2s_ABC}C). Heritability estimates for factors with simulated heritability of 0.0 were clustered near zero. However, if larger numbers of factors (scenarios \emph{b}-\emph{c}), or fewer traits (scenario \emph{f}) were simulated, more of these non-genetic factors were estimated to be have $h^2 > 0.05$ (Figure~\ref{fig:factor_h2s_ABC}). In scenario \emph{e}, the five simulated factors were all assigned a heritability of 1.0, but the residual covariance matrix $\mathbf{R}$ did not have a factor structure. Our model estimates these factors as having high heritability ($\sim0.9$, Figure~\ref{fig:factor_h2s_ABC}B). In scenarios \emph{h}-\emph{j}, simulated heritabilities of the five genetic factors were varied between 0.9 and 0.1 (Figure~\ref{fig:factor_h2s_D}). With moderate-large sample sizes (scenarios \emph{i}-\emph{j}), all factor heritability estimates were accurate, though some downward-bias was evident for the lower-heritability factors. With low sample sizes, factor heritability estimates were noisier, both for the genetic and non-genetic factors, and the downward bias was more apparent. Figure~\ref{fig:trait_h2s} shows the accuracy of the estimated trait heritabilities across the 20-1,000 traits in each scenario. Each datapoint represents the square root of the mean squared error of trait heritabilities fit for one of the 10 simulations of each scenario. Interestingly, the most accurate trait heritability estimates were recovered when $\mathbf{R}$ had a factor structure, but was not sparse (scenario \emph{d}, Figure~\ref{fig:trait_h2s}B). Heritability estimates were more accurate with increasing complexity of $\mathbf{G}$ and $\mathbf{R}$ (Figure~\ref{fig:trait_h2s}A), or increasing sample size (Figure~\ref{fig:trait_h2s}D). The average accuracy was not strongly affected by the number of traits studied (Figure~\ref{fig:trait_h2s}C), or the form of the residual covariance matrix (Figure~\ref{fig:trait_h2s}B).

\subsection{Gene expression example:} 
 
\begin{figure}[htbp!]
\begin{center}
\includegraphics[width=6in]{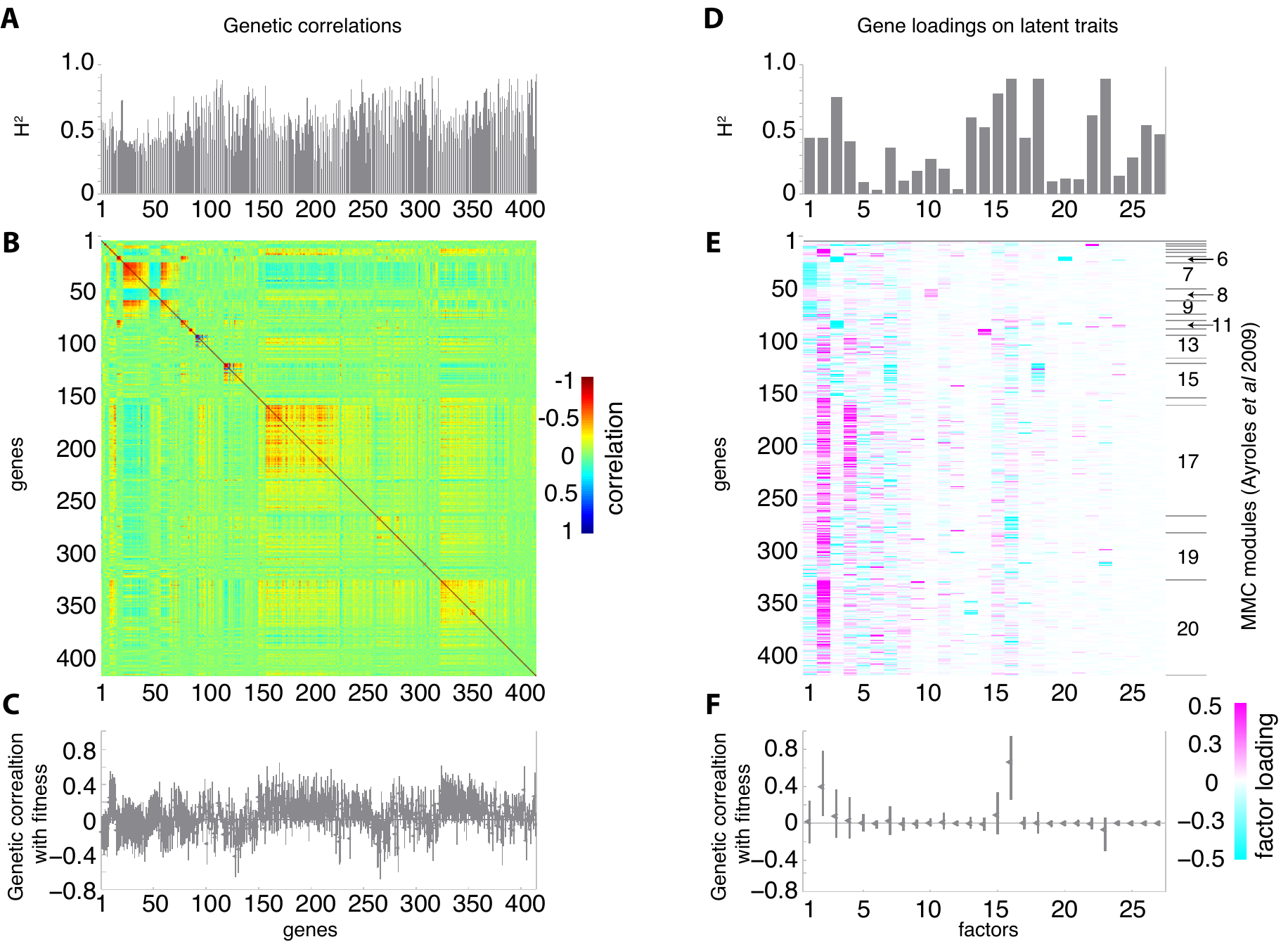}
\caption[Estimated G for Drosophila data]{ \textbf{Among-line covariance of gene expression and competitive fitness in Drosophila is modular.} \textbf{A-C} Genetic (among-line) architecture of 414 gene expression traits \cite{Ayroles:2009gd}. \textbf{A}. Posterior mean broad-sense heritabilities ($H^2$) for the 414 genes. \textbf{B}. Posterior mean genetic correlations among these genes. \textbf{C}. Posterior means and $95\%$ highest posterior density (HPD) intervals around estimates of genetic correlations between each gene and competitive fitness. For comparison, see Figure 7a of \citeN{Ayroles:2009gd}. \textbf{D-F}. Latent trait structure of gene expression covariances. \textbf{D} Posterior mean $H^2$ for each estimated latent trait. \textbf{E}. Posterior mean gene loadings on each latent trait. \textbf{F}. Posterior means and $95\%$ (HPD) intervals around estimates of genetic correlations between each latent trait and competitive fitness. The right-axis of panel \textbf{E}. groups genes into modules inferred using Modulated Modularity Clustering \cite{Stone:2009jx,Ayroles:2009gd}.}
\label{fig:Fitness_G}
\end{center}
\end{figure}

Our estimate of the G-matrix from the Drosophila gene expression data was qualitatively similar to the original estimate (Figure~\ref{fig:Fitness_G}B, and compare to Figure 7a in \citeN{Ayroles:2009gd}). Estimates of the broad-sense heritability of each gene were also similar ($r = 0.74$). While a direct comparison of the dominant G-matrix subspace recovered by our model and the estimate by \citeN{Ayroles:2009gd} was not possible because individual covariances were not reported, we could compare the two estimates of the underlying structure. Using the Modulated Modularity Clustering (MMC) algorithm \cite{Stone:2009jx}, \citeN{Ayroles:2009gd} identified 20 modules of genetically correlated transcripts \textit{post-hoc}. Our model identified 27 latent factors (Figure~\ref{fig:Fitness_G}D-F), of which 13 were large factors (explaining $> 1\%$ variation in $2^+$ genes). The large factors were consistent ($r > 0.95$) across three 3 parallel chains of the Gibbs sampler. Many factors were similar to the modules identified by MMC (Figure~\ref{fig:Fitness_G}E). Some of the factors were nearly one-to-one matches to modules (e.g., factor 10 with module 8, and factor 14 with module 12). However, others merged together two or more modules (e.g., factor 1 with modules 7 and 9, and factor 2 with modules 4, 13, 16-20). And some entire modules were part of two or more factors (e.g., module 17 was included in factors 2 and 4, and module 18 was included in factors 2 and 16).

Each factor (column of $\mathbf{\Lambda}$) represents a sparse set of genes that are highly correlated in their expression, possibly due to common regulation by some latent developmental trait. Using the Database for Annotation, Visualization and Integrated Discovery (DAVID) v6.7 \cite{Huang:2009be,Huang:2009gk}, we identified several factors that were individually enriched (within this set of 414 genes) for defense and immunity, nervous system function, odorant binding, and transcription and cuticle formation. Similar molecular functions were identified among the modules identified by \citeN{Ayroles:2009gd}. By inferring factors at the level of phenotypic variation, rather than the among-line covariances, we could directly estimate the broad-sense heritability ($H^2$) of these latent traits themselves. Figure~\ref{fig:Fitness_G}D shows these $H^2$ estimates for each latent trait. Several of the factors have very low ($<0.2$) or very high ($>0.75$) $H^2$ values. Selection on the later  latent traits would likely be considerably more efficient than the former.

Finally, by adding a competitive fitness as a 415th trait in the analysis, we could estimate the among-line correlation between the expression of each gene and this fitness-related trait (Figure~\ref{fig:Fitness_G}C). Many (60/414 $\sim 15\%$ of all genes analyzed) of the $95\%$ highest posterior density (HPD) intervals on the among-line correlations did not included zero, although most of these correlations were low (for $85\%$ of genes, $|r| < 0.25$) with a few as large as $|r| \sim 0.45$. More significantly, we could also estimate the genetic correlation between competitive fitness and each of the latent traits defined by the 27 factors (Figure~\ref{fig:Fitness_G}F). Most factors had near-zero genetic correlations with competitive fitness. However, the genetic correlations between competitive fitness and factors 2 and 16 were large and highly significant, suggesting potentially interesting genetic relationships between these two latent traits and fitness.

\section{Discussion}
	The Bayesian genetic sparse factor model performs well on both simulated and real data, and thus opens the possibility of incorporating high dimensional traits into evolutionary genetic studies and breeding programs. Technologies for high-dimensional phenotyping are becoming widely available in evolutionary biology and ecology so methods for modeling such traits are needed. Gene expression traits in particular provide a way to measure under-appreciated molecular and developmental traits that may be important for evolution, and technologies exist to measure these traits on very large scales. Our model can also be applied to other molecular traits (e.g., metabolites or protein concentrations), high dimensional morphological traits (e.g., outlines of surfaces from geometric morphometrics), or gene-environment interactions (e.g., the same trait measured in multiple environments).

\subsection{Scalability of the method:} The key advantage of the Bayesian genetic sparse factor model over existing methods is its ability to provide robust estimates of covariance parameters for datasets with large numbers of traits. In this study, we demonstrated high performance of the model for $100-1,000$ simulated traits, and robust results on real data with 415. Similar factor models (without the genetic component) have been applied to gene expression datasets with thousands of traits \cite{Bhattacharya:2011gh}, and we expect the genetic model to perform similarly. The main limitation will be computational time, which scales roughly linearly with the number of traits analyzed (assuming the number of important factors grows more slowly). As an example, analyses of simulations from scenario \emph{g} with 1,000 traits and 1,000 individual took about 4 hours to generate 12,000 posterior samples on a laptop computer with a 4-core 2.4 GHz Intel Core i7, while analyses of scenario \emph{a} with 100 traits took about 45 minutes. Parallel computing techniques may speed up analyses in cases of very large (e.g., 10,000+) numbers of traits. 

The main reason that our model scales well in this way is that under our prior, each factor is sparse. Experience with factor models in fields such as gene expression analysis, economics, finance, and social sciences \cite{Fan:2011fi}, as well as with genetic association studies (e.g., \citeNP{englehardt,Stegle:2010cg,Parts:2011df}) demonstrates that sparsity (or shrinkage) is necessary to perform robust inference on high-dimensional data \cite{bickel1,bickel2,elkaroui,Meyer:2010jj}. Otherwise, sampling variability can overwhelm any true signals, leading to unstable estimates. Here, we used the \textit{t}-distribution as a shrinkage prior, following \cite{Bhattacharya:2011gh}, but many other choices are possible \cite{Armagan:2011tw}.

\subsection{Applications to evolutionary quantitive genetics:} The G-matrix features prominently in the theory of evolutionary quantitative genetics, and its estimation has been a central goal of many experimental and observational studies \cite{Walsh:2009gg}. Since our model is built on the standard ``animal model" mixed effect model framework, it is flexible and can be applied to many experimental designs or studies. And since our model is Bayesian and naturally produces estimates within the parameter space, posterior samples from the Gibbs sampler provide convenient credible intervals for the G-matrix itself and many evolutionarily important parameters, such as trait-specific heritabilities or individual breeding values \cite{Sorensen:2010wf}.  

An important use of G-matrices is to predict the response of a set of traits to selection \cite{Lande:1979th}. Applying Robertson's 2nd theorem of natural selection, the response in $\mathbf{\bar{y}}$ will equal the additive genetic covariance between the vector of traits and fitness ($\Delta \mathbf{\bar y} = \sigma_A(\mathbf{y},\bar{w})$) \cite{RAUSHER:1992vj,Walsh:2009gg}. This quantity can be estimated directly from our model if fitness is included as the $p^*= (p+1)$th trait:
$$\Delta \mathbf{\bar y} = \mathbf{\Lambda}_{u_{/p^*}} \mathbf{\Lambda}^T_{u_{p^*}},$$
\noindent where $\mathbf{\Lambda}_{u_{/p^*}}$ contains all rows of $\mathbf{\Lambda}_u$ except the row for fitness, and $\mathbf{\Lambda}_{u_{p^*}}$ contains only the row of $\mathbf{\Lambda}_u$ corresponding to fitness. Similarly, the quantity $1-\Psi_{u_{p^*}} / \mathbf{G}_{p^*,p^*}$ equals the percentage of genetic variation in fitness accounted for by variation in the observed traits \cite{Walsh:2009gg}, which is useful for identifying other traits that might be relevant for fitness.

On the other hand, our model is not well suited to estimating the dimensionality of the G-matrix. A low-rank G-matrix means that there are absolute genetic constraints on evolution \cite{Lande:1979th}. Several methods provide statistical tests for the rank of the G-matrix (e.g., \citeNP{Hine:2006ki,Kirkpatrick:2004bv,Mezey:2005wp}). We use a prior that shrinks the magnitudes of higher index factors to provide robust estimates of the largest factors. This will likely have a side-effect of underestimating the total number of factors, although this effect was not observed in our simulations. However, absolute constraints appear rare \cite{Houle:2010jw}, and the dimensions of the G-matrix with the most variation are likely those with the greatest effect on evolution in natural populations \cite{Schluter:1996up,Kirkpatrick:2009er}. Our model should estimate these dimensions well. From a practical standpoint, pre-selecting the number of factors has plagued other reduced-rank estimators of the G-matrix (e.g., \citeNP{Kirkpatrick:2004bv,Hine:2006ki,Meyer:2009jm}).
Our prior is based on an infinite factor model \cite{Bhattacharya:2011gh}, and so no \textit{a priori} decision is needed. Instead, the parameters of the prior distribution become important modeling decisions. In our experience, a relatively diffuse prior on $\delta_l$ with $a_2=3, b_2=1$ tends to work well. 

\subsection{Biological interpretation of factors:}	 
Genetic modules are sets of traits likely to evolve together. By assuming that the developmental process is modular, we can model each latent trait as affecting a limited number of observed traits. A unique feature of our model is the fact that we estimate genetic and environmental factors jointly, instead of separately as in classic multilevel factor models (e.g., \citeNP{goldstein118}). If each factor represents a true latent trait (e.g., variation in a developmental process), it is reasonable to decompose variation in this trait into genetic and environmental components. We directly estimate the heritability of the traits underlying each factor, and therefore can use our model to predict the evolution of these latent traits.

Other techniques for identifying genetic modules have several limitations. The MMC algorithm \cite{Stone:2009jx,Ayroles:2009gd} does not infer modules in an explicit quantitative genetic framework, and constraints each trait to belong to only one module. In some analyses (e.g., \citeNP{Mcgraw:2011dm}), each major eigenvector of the $\mathbf{G}$ or $\mathbf{P}$ matrices is treated as underlying module. These eigenvectors can be modeled directly (e.g., \citeNP{Kirkpatrick:2004bv}), but the biological interpretation of the eigenvectors is unclear because of the mathematical constraint that the they be orthogonal \cite{Hansen:2008by}. In classic factor models (such as proposed by \citeN{Meyer:2009jm}, or \citeN{deLosCampos:2007ib}), factors are not identifiable \cite{Meyer:2009jm}, and so the identity of the underlying modules is unclear. Under our sparsity prior, factors are identifiable (up to a sign-flip: the loadings on each factor can be multiplied by $-1$ without affecting its probability under the model, but this does not change which traits are associated with each factor). In simulations and with the Drosophila gene expression data, independent MCMC chains consistently identify the same dominant factors. Therefore the observed traits associated with each factor can reliably be used to characterize the developmental process represented by the latent trait.

\subsection{Extensions:}
Our model is built on the classic mixed effect model common in quantitative genetics \cite{Henderson:1984vh}. It is therefore straightforward to extend to models with additional fixed or random effects (e.g., dominance or epistatic effects) for each trait. However, the update equation for $h^2_j$ in the Gibbs sampler described in the Appendix does not allow additional random effects in the model for the latent factors themselves, although other formulations are possible. A second extension relates to the case when the relationship matrix among individuals ($\mathbf{A}$) is unknown. Here, relationship estimates from genotype data can be easily incorporated. As such, our model is related to a recently proposed sparse factor model for genetic associations with intermediate phenotypes \cite{Parts:2011df}. These authors introduced prior information on genetic modules from gene function and pathway databases which could be incorporated in our model in a similar way.

\section{Conclusions}
The Bayesian genetic sparse factor model for genetic analysis that we propose provides a novel approach to genetic estimation with high-dimensional traits. We anticipate that incorporating many diverse phenotypes into genetic studies will provide powerful insights into evolutionary processes. The use of highly-informative but biologically grounded priors is necessary for making inferences on high-dimensional data, and can help identify developmental mechanisms underlying phenotypic variation in populations.

\section{Appendix}
\subsection{Posterior sampling:}\label{postsamp}
We estimate the posterior distribution of the Bayesian genetic sparse
factor model with an adaptive Gibbs sampler based on the procedure proposed 
by \citeN{Bhattacharya:2011gh}. The value $k^*$ at
which columns in $\mathbf{\Lambda}$ are truncated is set using
an adaptive procedure \cite{Bhattacharya:2011gh}.
Given a truncation point, the sampler iterates through the following steps:

\begin{enumerate}

\item If missing observations are present, values are drawn independently from univariate normal distributions parameterized by the current values of all other parameters:
\begin{align*}
\pi(y_{im}\mid - ) \sim \mbox{N}\left(\mathbf{x}^{(j)}\mathbf{b}_i + \mathbf{f}^{(m)}\bm{\lambda}_i + \mathbf{z}^{(m)}\bm{\delta}_i,(\sigma^{-2}_i)^{-1}\right)
\end{align*} 
\noindent where $y_{im}$ is the imputed phenotype value for the $i$-th trait in individual $m$. The three components of the mean are: $\mathbf{x}^{(m)}$ the row vector of fixed effect covariates for individual $m$ times $\mathbf{b}_i$, the $i$th column of the fixed effect coefficient matrix; $\mathbf{f}^{(m)}$, the row vector of factor scores on the $k^*$ factors for individual $m$ times  $\bm \lambda_i$, the row of the factor loading matrix for trait $i$; and $\mathbf{z}^{(m)}$, the row vector of the random (genetic) effect incidence matrix for individual $m$ times  $\bm{\delta}_i$, the vector of residual genetic effects for trait $i$ not accounted for by the $k^*$ factors. Finally, $\sigma^{-2}_i$ is the residual precision of trait $i$. All missing data can be drawn in a single block update.

\item The fixed effect coefficient matrix $\mathbf{B}$, the truncated factor loading matrix $\mathbf{\Lambda}_{k^*}$ and the residual genetic effects matrix $\mathbf{\Delta}$ can be stacked into a single matrix, and then its columns factor into independent multivariate normal conditional posteriors:

\begin{align*}
\pi\left(\left[\begin{array}{c}
  \mathbf{b}_i \\ 
  \bm \delta_i \\ 
  \bm \lambda_i \\ 
\end{array}\right] \big| - \right) &\sim \mbox{N}\left(\mathbf{C}^{-1}\mathbf{W}^T\sigma_i^2\mathbf{y}_i,\mathbf{C}^{-1}\right),
\end{align*}
\noindent where $\mathbf{W}$ and $\mathbf{C}$ are defined as:
\begin{align*}
\mathbf{W} &= [\mathbf{X}\; \mathbf{Z} \; \mathbf{F}] \\
\mathbf{C} &= 
\left[\begin{array}{ccc}
 \mathbf{0} &  \mathbf{0}  &  \mathbf{0} \\
  \mathbf{0}  & \psi_{a_{ii}}^{-2} \mathbf{A}^{-1} &  \mathbf{0}  \\
   \mathbf{0} &  \mathbf{0}  & \mbox{Diag}(\phi_{ij}\tau_j)
\end{array}\right]  + \sigma_i^{-2}\mathbf{W}\mathbf{W}^T.
\end{align*}

\item The conditional posterior of the factor scores $\mathbf{F}$ is a matrix variate normal distribution:
\begin{align*}
\pi\left(\mathbf{F} \mid - \right) &\sim \mbox{MN}_{n,k^*}\left(\mathbf{C}^{-1}\left(\tilde{\mathbf{Y}}\mathbf{\Psi}_e^{-1}\mathbf{\Lambda_{k^*}} + \mathbf{Z}\mathbf{F}_u\mbox{Diag}(1-h^2_i)^{-1}\right),\mathbf{C}^{-1}\right)
\end{align*}
\noindent where $\mathbf{C}$ is:
\begin{align*}
\mathbf{C} &= \mathbf{\Lambda}_{k^*}^T \mathbf{\Psi}_e^{-1} \mathbf{\Lambda}_{k^*} +\mbox{Diag}(1-h^2_i)^{-1}
\end{align*}
\noindent and $\tilde{\mathbf{Y}}$ is:
\begin{align*}
\tilde{ \mathbf{Y}} = \mathbf{Y}-\mathbf{X}\mathbf{B} - \mathbf{Z}\bm \Delta.
\end{align*}

\item The conditional posterior of the genetic effects on the factors, $\mathbf{F}_u$ factors into independent multivariate normals for each factor $\mathbf{f}_{u_j}, j = 1 \dots k^* \mbox{ st } h^2_j \neq 0$:
\begin{align*}
\pi\left(\mathbf{f}_{u_j} \mid - \right) &\sim \mbox{MN}\left(\mathbf{C}^{-1}(1-h^2_j)^{-1}\mathbf{Z}\mathbf{F}_m,\mathbf{C}^{-1}\right)
\end{align*}
\noindent where $\mathbf{C}$ is:
\begin{align*}
\mathbf{C} &= (1-h^2_j)^{-1}\mathbf{Z}\mathbf{Z}^T + (h^2_j)^{-1}\mathbf{A}^{-1}.
\end{align*}

\item The conditional posterior for each of the latent factor heritabilities $h^2_j, j = 1 \dots k^* $ is calculated by integrating out $\mathbf{F}_u$ and summing over all possibilities of $h^2_j$, since the prior on this parameter is discrete:
\begin{align*}
\pi\left(h^2_j = h^2 \mid - \right) &= \frac{\mbox{N}\left(\mathbf{F}_j \mid \mathbf{0}, h^2\mathbf{Z}\mathbf{A}\mathbf{Z}^T + (1-h^2) \mathbf{I}_n\right) \pi(h^2_j=h^2)}{\sum\limits_{l=1}^{n_h}\mbox{N}\left(\mathbf{F}_j \mid \mathbf{0}, h^2_l\mathbf{Z}\mathbf{A}\mathbf{Z}^T + (1-h^2_l) \mathbf{I}_n\right) \pi(h^2_j=h^2_l)}
\end{align*}
\noindent where $\mbox{N}(\mathbf{x}\mid \bm \mu,\Sigma)$ is the multivariate normal density with mean $\bm \mu$ and variance $\Sigma$, evaluated at $\mathbf{x}$, $h^2_l = l/n_h$, and $\pi(h^2_j=h^2)$ Is the prior probability that $h^2_j = h^2$. Given this conditional posterior, $h^2_j$ is sampled from a multinomial distribution.

\item The conditional posterior of the trait-factor loading variance $\phi_{ih}$ for trait $i$ on factor $h$ is:
\begin{align*}
\pi(\phi_{ih}\mid -) \sim \mbox{Ga}\left(\frac{\nu+1}{2},\frac{\nu + \lambda_{ih}^2}{2}\right).
\end{align*}

\item The conditional posterior of $\delta_m, m=1\dots k^*$ is as follows. For $\delta_1$:
\begin{align*}
\pi(\delta_1\mid -) \sim \mbox{Ga}\left(a_1+\frac{pk^*}{2},b_1+\frac{1}{2}\sum\limits_{l=1}^{k^*}\tau_l^{(1)}\sum\limits_{j=1}^p \phi_{jl}\lambda^2_{jl}\right)
\end{align*}
\noindent and for $\delta_h, h\geq 2$:
\begin{align*}
\pi(\delta_h\mid -) \sim \mbox{Ga}\left(a_2+\frac{p}{2}(k^*-h+1),b_2+\frac{1}{2}\sum\limits_{l=h}^{k^*}\tau_l^{(h)}\sum\limits_{j=1}^p \phi_{jl}\lambda^2_{jl}\right)
\end{align*}
\noindent where $\tau_l^{(h)} = \prod\limits_{t=1,t\neq h}^l\delta_t$ for $h=1\dots k^*$.

\item The conditional posteriors for the precision of the residual genetic effects of trait $i$, $\psi_{u_{ii}}$, is:
\begin{align*}
\pi(\psi_{u_{ii}}\mid -) \sim \mbox{Ga}\left(a_g + \frac{r}{2},b_g + \frac{1}{2}\bm \delta_i^T \bm \delta_i\right).
\end{align*}

\item The conditional posteriors for the model residuals of trait $i$, $\sigma_i^{-2}$, is:
\begin{align*}
\pi(\sigma_i^{-2}\mid -) \sim \mbox{Ga}\left(a_r + \frac{n}{2},b_r + \frac{1}{2}\sum\limits_{j=1}^n \left(y_{ij}-\mathbf{x}^{(j)}\mathbf{b}_i - \mathbf{f}^{(j)}\bm{\lambda}_i - \mathbf{z}^{(j)}\bm{\delta}_i\right)^2\right).
\end{align*}

\end{enumerate}

Other random effects, such as the line $\times$ sex effects modeled in the gene expression example of this paper can be incorporated into this sampling scheme in much the same way as the residual genetic effects, $\mathbf{\Delta}$, are included here.

\section{Acknowledgments}
	We would like to thank Barbara Engelhardt, Iulian Pruteanu-Malinici, Jenny Tung, and two anonymous reviewers for comments and advice on this method.

\bibliography{paper_library3}
\bibliographystyle{mychicago}

\end{document}